# Accurate high speed single-electron quantum dot preparation

S. P. Giblin[1*], S. J. Wright[2,3], J. Fletcher[1], M. Kataoka[1], M. Pepper[4], T. J. B. M. Janssen[1], D. A. Ritchie[2], C. A. Nicoll[2], D. Anderson[2] and G. A. C. Jones[2].

[1]National Physical Laboratory, Hampton Road, Teddington, Middlesex TW11 0LW, UK
[2]Cavendish Laboratory, University of Cambridge, J. J. Thomson Avenue, Cambridge CB3 0HE, UK
[3]Toshiba Research Europe Ltd, Cambridge Research Laboratory, 208 Science Park, Milton Road, Cambridge CB4 0WE, UK.
[4]University College London, Torrington Place, London WC1E 7JE, UK

**Abstract.** Using standard microfabrication techniques it is now possible to construct devices, which appear to reliably manipulate electrons one at a time. These devices have potential use as building blocks in quantum computing devices, or as a standard of electrical current derived only from a frequency and the fundamental charge. To date the error rate in semiconductor 'tuneable-barrier' pump devices, those which show most promise for high frequency operation, have not been tested in detail. We present high accuracy measurements of the current from an etched GaAs quantum dot pump, operated at zero source-drain bias voltage with a single AC-modulated gate driving the pump cycle. By comparison with a reference current derived from primary standards, we show that the electron transfer accuracy is better than 15 parts per million. High-resolution studies of the dependence of the pump current on the quantum dot tuning parameters also reveal possible deviations from a model used to describe the pumping cycle.

Devices that can reliably transfer electrons one at a time, electron pumps, have important applications in the fields of electrical metrology [1], and solid-state quantum computing [2,3]. In the former field, there is especially strong interest, motivated by re-defining the SI base unit Ampere in terms of the electron charge and a known frequency [4,5]. Pumps based on multiple metal-oxide tunnel junctions have demonstrated very high relative accuracies approaching $10^{-8}$ [6], but the speed of transfer is limited to around 10 MHz by the intrinsic time-constant of the junctions. The resulting pumped current $\approx 1$ pA is at least an order of magnitude too small for the pump to function as a useful current standard, although it was used to demonstrate a quantum capacitance standard by charging a capacitor with a known number of electrons [7]. More recently, an innovative device, the "hybrid turnstile" has been demonstrated, utilizing metal-oxide-superconductor tunnel junctions [8]. Unlike the multiple-junction pumps, the hybrid turnstile needs only one AC control signal, and consequently the current can be increased by operating many devices in parallel [9]. However, the hybrid turnstile needs to be operated at finite bias voltage $\approx 1$ mV, and eliminating errors due to leakage currents is a challenging ongoing project [10].

Another class of electron pumps exploits the tunability of potential barriers in reduced-dimensional semiconductor systems. Following the pioneering work of Kouwenhoven *et al*, on pumping electrons through a quantum dot at finite source-drain bias [11], it was found that electrons could be pumped through a dot at *zero* source-drain bias by applying a large AC modulation to just *one* of the gates [12,13], as illustrated schematically in fig. 1a for the simplest case of one electron pumped for each cycle. The experimental signature of pumping is a DC current $I_\mathrm{P} \approx I_0 \equiv n_0 ef$, where $f$ is the repetition frequency of the potential modulation, and $n_0$ is an integer. Pumping has been observed in etched GaAs 2-dimensional electron gas (2-DEG) quantum dots [12,14] and silicon nano-wire MOSFETs [13], for $f$ up to the order of 1 GHz. Furthermore, parallel operation of two GaAs pumps has recently been demonstrated

[15]. The high operation speed, zero source-drain bias, and possibility of parallel current scaling, makes these pumps promising candidates for a primary metrological current source [1], as well as a source of single electrons for semiconductor-based quantum logic gates [3]. A crucial unanswered question addressed in this work concerns the accuracy of the electron transfer in the semiconductor pump. Estimates of acceptable error rates for fault-tolerant quantum computing range from 1 in $10^2$ to $10^6$ qubit operations [16], while metrological application of electron pumps as quantum standards of current requires error rates less than 1 in $10^7$ [4,5]. In contrast, normal laboratory measurements of the fractional error in the pumped current $\Delta I_P = (I_P - I_0)/I_0$, have at best shown $|\Delta I_P| \leq 10^{-2}$. One study, using a calibrated ammeter to measure $I_P$, set a lower limit to possible errors: $|\Delta I_P| \leq 10^{-4}$, but this level of accuracy was only observed over a very narrow range of gate voltages used to tune the pump operating point [14]. Theoretical treatments of the error rates in these pumps are much more difficult than for the case of low-frequency fixed-barrier pumps [17,18], partly due to the rapid non-adiabatic change in the tunnel coupling from the source reservoir to the dot [19]. Some features of the pump behaviour have been explained by considering the time dependence of the back-tunneling rates during the initial phase of the pump cycle (Fig. 1a, frame 2) [12,13, 20,21], and predictions of $\Delta I_P \leq 10^{-5}$ have been made using this approach [21]. Furthermore, recent experiments showed that application of a magnetic field of a few Tesla results in an improvement in plateau flatness in GaAs pumps [22,23]. This suggests that spin states within the dot [24] or edge states in the leads [25] may play a role in the transport. With this promising experimental and theoretical background, high-accuracy measurements of the pump current are clearly of great interest. In this work we compare the current from GaAs pumps with a reference current derived from primary electrical standards with relative accuracy approaching $10^{-5}$. This enables us to set much more stringent limits on error mechanisms than was possible from previously measured data.

Our pumps were fabricated by wet-chemical etching of sub-micron width wires in a GaAs 2-DEG system, followed by deposition and patterning of Ti/Au surface gates [22]. An SEM image of a pump is shown in Fig. 1c, together with the circuitry for biasing the gates and measuring the current. The key feature of our measurement setup is the reference current $I_R$, with opposite polarity to $I_P$, which we generated by applying a linear voltage ramp to a low-loss capacitor [26]: $I_R = C dV_{CAL}/dt$. $I_R$ was traceable to primary maintained standards of capacitance, voltage and time, and had a relative systematic uncertainty of 15 parts per million (ppm). The magnitude of $I_R$ was adjusted to be within 0.2% of $I_P$; consequently the ammeter current $I = I_P - I_R$ was small and variations in the ammeter gain (for example due to ambient temperature changes) had a negligible affect on the result. To remove offsets in the measurement circuitry and reference current source, both the pump and reference currents were switched on and off with a cycle period of 60 s, and the pump current calculated from the difference signal. The raw ammeter readings from one pump cycle are shown in Fig. 1d. A sine wave at frequency $f = 340$ MHz applied to one gate implemented the pumping cycle, illustrated schematically in fig. 1 a and b. The pumps were mounted in a dilution refrigerator, and all data was taken at a mixing-chamber temperature of $\approx 30$ mK and a perpendicular magnetic field of 5 T. We investigated the pump behaviour as a function of four adjustable control parameters: the DC voltages applied to the two gates, $V_{GS}$ and $V_{GD}$, the RF generator power $P_{RF}$, and the source-drain bias voltage $V_B$. In all data apart from Fig. 3c, $V_B = 0$. Following each cool-down, the parameters were tuned iteratively to yield maximally flat quantised plateaus before making the measurements presented in this work.

Fig. 2a shows conventional low-resolution measurements of the pumped current as a function of the fixed gate voltage $V_{GD}$, for two samples denoted A and B. Both samples exhibit a wide plateau region over which $I_P = ef$ (horizontal black line) on the coarse scale of this graph, similar to previously reported results for GaAs pumps in a magnetic field [15,22,23]. We fitted the $I_P(V_{GD})$ data to a back-tunneling model [21], (solid lines in the plot), and obtained minimum values of the error from the fits of $\Delta I_{P,MIN} = -4\times 10^{-6}$ and $-1\times 10^{-5}$ for samples A and B respectively. The minimum error is obtained at the value of $V_{GD}$ for which $dI_P/dV_{GD}$ is minimum. Next we used our high-resolution measurement technique to zoom in on the plateau region. Fig 2b shows the result of two measurement runs, taken 24 hours apart on sample A, plotted on a current axis expanded by a factor of 5000 relative to fig. 2a. The ±15 ppm systematic uncertainty is indicated by a grey shaded region centred on $\Delta I_P=0$. The main conclusion of this study is apparent from this data: On the plateau region, there is no statistically significant offset of the pump current from $I_0$. The offset of ≈0.8 fA below $I_0$ is within the systematic uncertainty of the measurement system, and there is good agreement between the two measurement runs. The value of $\Delta I_{P,MIN}=-4\times 10^{-6}$ predicted from the fit A1 is consistent with our data: a weighted average of the two closest data points to the grey arrow in fig. 2b gives $\Delta I_{P,MIN}=(-5\pm 18)\times 10^{-6}$. However, it is clear that the fit to the low-resolution data (solid black line) considerably under-estimates the flatness of the plateau. The discrepancy may be evidence that the treatment of the back-tunnel rates in ref. 21 is over-simplified. An alternative explanation is that experimental artefacts such as noise pickup or a rectification process [27] leads to an apparent broadening of the transition between plateaus in the data of fig. 2a.

In a truly quantized system, the parameter of interest (pumped current in our case) should be invariant over a finite range of all the adjustable parameters. To investigate this, we measured the current as a function of $V_{GS}$, $V_{GD}$, $V_B$ and $P_{RF}$ for sample B, shown in fig. 3a-d. Our experimental wiring contributed 10 TΩ of leakage resistance in parallel with the pump, which can be resolved as a finite gradient $dI_P/dV_B$. This leakage does not constitute a significant source of error when the pump is operated close to zero $V_B$. Otherwise, each scan shows a plateau region flat to within the typical ≈10 ppm error in the slope of a linear fit. Fig. 3e shows average values of $\Delta I_P$ for all 6 high-resolution data sets presented in this paper, with shading to indicate the systematic uncertainty at 68% confidence (dark grey) and 95% confidence (light grey) intervals. The data points in this figure are the weighted averages of four consecutive points from each data set, chosen from the centre of each plateau (indicated by vertical lines in fig3 a-d). The difference between the mean currents for samples A and B is $\Delta I_P(A) - \Delta I_P(B) = (-5\pm2.5)\times10^{-6}$, indicating possible sample-dependent errors at the ppm level which will be investigated more closely in future work. Note that the comparison of the currents from two samples is limited only by the random uncertainty. A weighted average of all data points in the figure yields an overall estimate of the pump error, $\Delta I_P=(-14.8\pm15)\times10^{-6}$. Because we operate the pump in a regime where all the electrons loaded into the dot are ejected [28], we can also interpret our result as probing the reliability of loading the dot with just one electron in repeated operations: n=0.9999852±0.000015. This data is convincing evidence that the electron transport in tunable-barrier pumps is robustly quantised at the $10^{-5}$ level or better, over a useful range of parameter space.

The accuracy in our experiment is close to the limit of what can be achieved with conventional room-temperature instrumentation. To further improve the systematic uncertainty, a cryogenic current comparator (CCC) could be used to compare the pump and reference currents [29]. This would enable a direct test of the error rates of a few ppm predicted by the model in ref. 21, and place experimental limits on other types of error such as thermally activated tunnelling, which have been predicted to be negligible [20]. Furthermore, if independent confirmation of the pump transport accuracy could be obtained, for example by a shuttle-type experiment with an on-chip charge detector to detect individual transport errors [6], the same experimental setup would constitute realisation of the metrological triangle [5] which is one of the long-standing goals of fundamental metrology.

The authors would like to thank Bernd Kaestner and Chris Ford for stimulating discussions. The work was funded by the European Community's 7'th Framework Programme (ERA-NET

Plus Grant No. 217257), the EPSRC, and the UK Department for Business, Innovation and Skills. SJW acknowledges additional support from Toshiba Research Europe Ltd.

**Author Contributions**

S. P. G. designed the measurement system and wrote the paper. S. J. W. designed and fabricated the samples. S. P. G. and S. J. W. jointly performed the experiments and contributed equally to the overall work. J. F. did data analysis, and both J. F. and M. K. contributed conceptual and theoretical knowledge. M. P. and T. J. B. M. J. provided project leadership and supervision. The GaAs wafers were grown by D. A. R. & C. A. N., and D. A. & G. A. C. T did the electron beam patterning.

**Competing financial interest statement**

The authors declare no competing financial interests

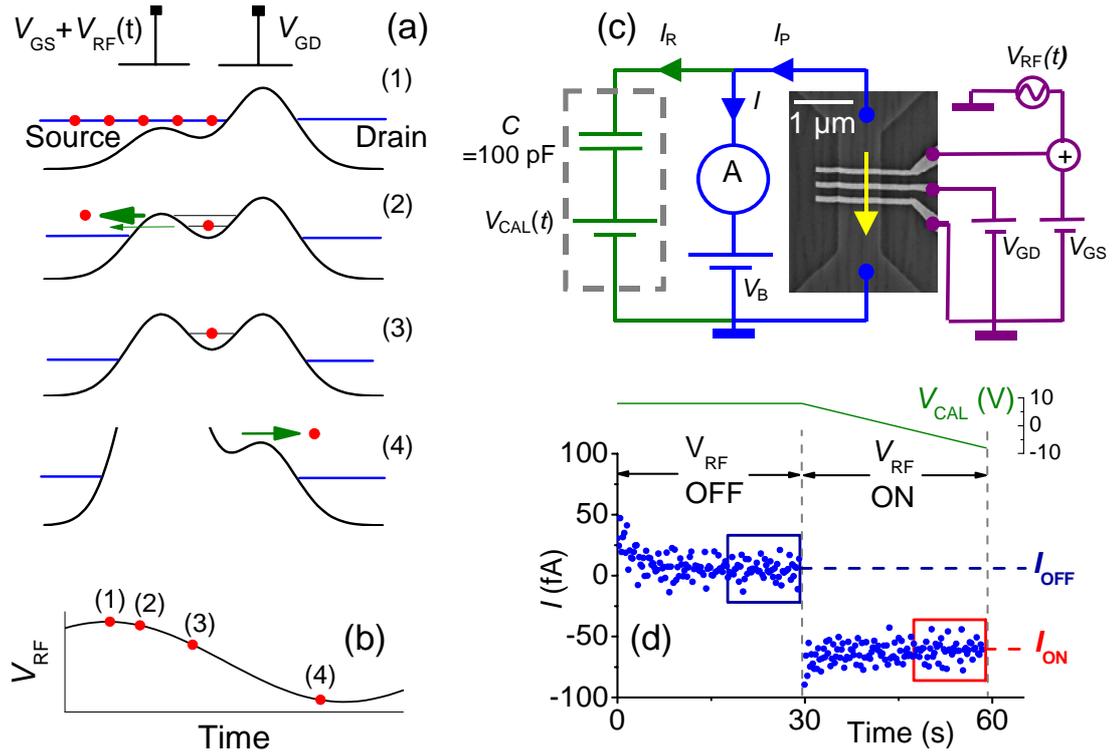

**Fig. 1. The pumping mechanism, measurement circuit and raw data. a**. Schematic energy diagram of the quantum dot illustrating the pumping cycle. A single electron is pumped from the source to the drain by modulating the left (source-side) potential barrier. Frames (1)-(4) show successive stages in the cycle. The green arrows in frame 2 indicate the back-tunneling of the second electron, and the subsequent small probability of the first electron back-tunneling, which would constitute a pumping error. **b:** Illustration of the points in the RF cycle corresponding to frames (1)-(4) in fig. 1 a. **c:** Schematic diagram of the measurement (blue), reference current (green) and gate bias (purple) circuitry, incorporating an SEM image of a device similar to the ones studied. The conducting channel, running from top to bottom, appears dark grey, and the metallic gates are the bright fingers. The lowermost gate, not used in this experiment, is grounded and the dot is formed between the top and middle gates. The yellow arrow indicates the direction of electron pumping. **d**. Main panel: a section of raw measured data. During the "off" phase, the RF source is turned off, and $V_{CAL}(t)$, plotted in the small upper panel, is held at a constant value. In the "on" phase, the RF source is on, and $V_{CAL}$ is ramped at $dV_{CAL}/dt \approx -0.5447$ V/s. Approximately 2/3 of the data from each phase is discarded, to allow for the time constant of the ammeter. The blue and red coloured boxes show the range of data points used to calculate average values, $I_{OFF}$ and $I_{ON}$ respectively.

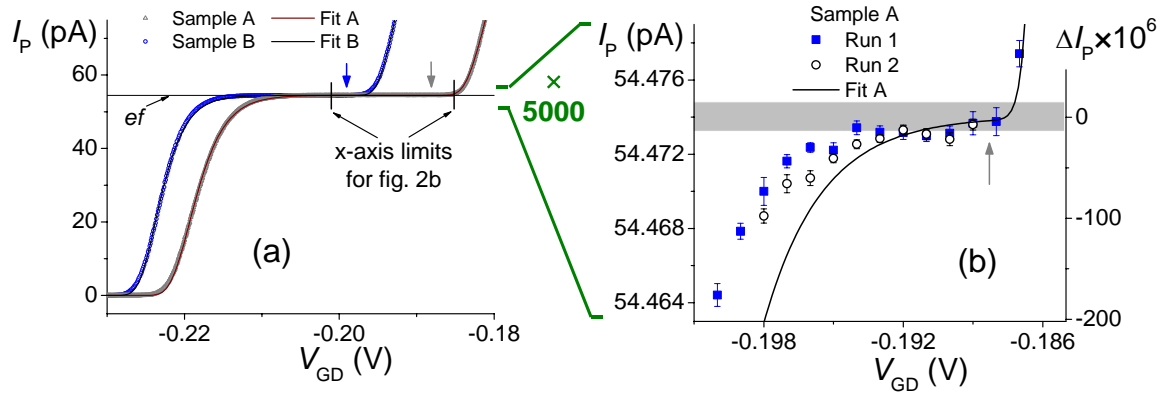

**Fig. 2 Pumped current as a function of the drain-side gate voltage. a**. Measurement over a wide range of $V_{GD}$ showing the 1-electron plateau, for two samples pumped at $f = 340$ MHz. For these low-resolution measurements, the reference current source was not used: $I = I_P$. Fits to the model of ref. 21 are shown as solid lines. The two vertical tick marks show the range of $V_{GD}$ investigated in the high-resolution data set of Fig. 2b. Vertical arrows indicate the values of $V_{GD}$ at which the derivative of the fit lines is minimum for fit A (grey arrow) and fit B (blue arrow). **b**. High-resolution measurement of the pump current for sample A, showing data for two measurement runs. Each data point is the average of many ON-OFF cycles illustrated in Fig. 1d. The error bars on the data points indicate the random uncertainty, and a shaded grey area around $I_0 \equiv ef = 54.47400$ pA shows the $\pm 15$ part per million 1 $\sigma$ systematic uncertainty. Fit A obtained from the data of fig. 2a is shown as a solid black line. The grey arrow has the same meaning as in fig. 2a.

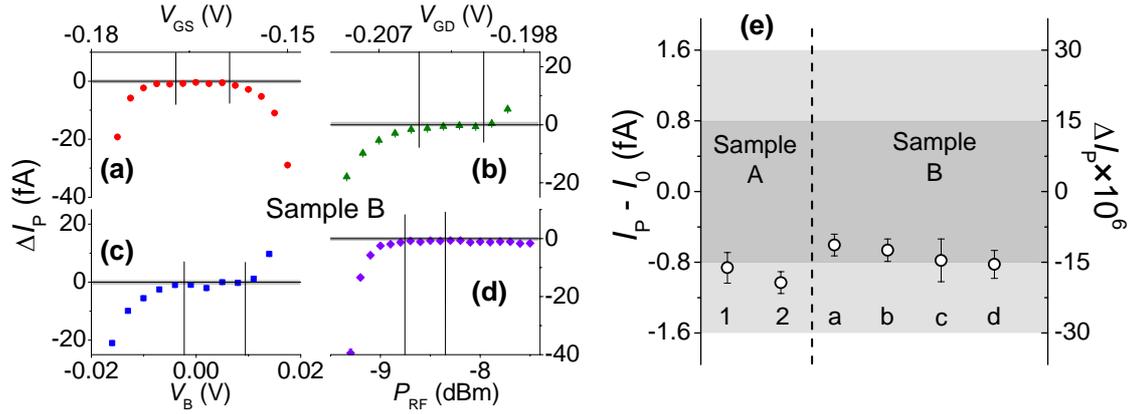

**Fig. 3 Pumped current as a function of all control parameters, and averaged current. a-d**: pump current relative to $I_0 \equiv ef = 54.47400$ pA for sample B, as a function of four control parameters $V_{GS}$, $V_{GD}$, $V_B$ and $P_{RF}$. The vertical lines indicate the group of 4 adjacent data points with minimal slope. The weighted mean of these points is plotted in fig. 3e. **e**. Mean pump current on the quantised plateau, calculated as a weighted average of the four data points with minimum gradient, for the six high-resolution data sets in figs 2b and 3a-d. The shaded regions show the systematic uncertainty at 68% confidence (dark grey) and 95 % confidence (light grey). The two data points for sample A correspond to the two experimental runs plotted in fig. 2b.